\documentclass[aps,prb,showpacs,amsmath,amssymb,twocolumn]{revtex4}

\usepackage{graphicx}

\usepackage{bm}
\usepackage{gensymb}

\begin{document}

\title{Structural evolution in high-pressure amorphous CO$_2$ from \textit{ab initio} molecular dynamics}

\author{Du\v{s}an Pla\v{s}ienka}
\email{plasienka@fmph.uniba.sk}
\affiliation{Department of Experimental Physics, Comenius University,
\\ Mlynsk\'{a} Dolina F2, 842 48 Bratislava, Slovakia}

\author{Roman Marto\v{n}\'{a}k}

\affiliation{Department of Experimental Physics, Comenius University,
\\ Mlynsk\'{a} Dolina F2, 842 48 Bratislava, Slovakia}


\date{\today}

\begin{abstract}

By employing \textit{ab initio} molecular dynamics simulations at
constant pressure we investigated behavior of amorphous carbon dioxide
between 0-100 GPa and 200-500 K. We focused on evolution of the
high-pressure polymeric amorphous form known as a-carbonia on its way
down to zero pressure, where it eventually converts into a molecular
state. During the simulations we observed a spectrum of amorphous
forms between two limiting polymeric forms with different proportion
of three and four-coordinated carbon atoms. Besides that we also found
a new mixed molecular-polymeric form that shows pronounced metastability
at certain conditions. The observed behavior suggests CO$_2$ as possible
candidate for polyamorphism. We discuss structural and physical properties
of the observed amorphous forms as well as their relations to crystalline
phases.

\end{abstract}

\maketitle

\section{Introduction}

Pressure-induced amorphization (PIA) and amorphous-amorphous transition (AAT), as well as liquid-liquid transition (LLT), are fundamentally interesting and widely studied phenomena occurring in some common materials \cite{Sharma-Sikka-PIA-review, McMillan-AAT-review-1, WWMcMillan-AAT-review}. Especially interesting is the existence of polyamorphism, both in the solid regime - AAT, as observed in H$_2$O, SiO$_2$, GeO$_2$ \cite{Itie-GeO2, Guthrie-GeO2, Hong-GeO2, Lelong-GeO2}, Si \cite{Deb-Si}, Ge \cite{Principi-Ge}, S \cite{Sanloup} or C \cite{Lin-C} and in the liquid state - LLT, as reported experimentally in P \cite{Katayama-LLT-P} and S \cite{Bellissent} (and disputed in N \cite{Mukherjee-Boehler-LLT-N, Goncharov-LLT-N, Boates-Bonev-LLT-N-1}) or predicted theoretically for C \cite{Glosli-Ree} and H \cite{Scandolo-H-LLT}. Recently, carbon dioxide has been found to enrich this class of materials for observing the AAT between extended a-carbonia and molecular amorphous forms \cite{Santoro-1, Kume} and also for prediction of LLT between molecular and polymeric liquids \cite{Boates-CO2-LLT-1, Boates-CO2-LLT-2}.

Carbon dioxide is one of the most important compounds found on Earth and in the Solar system, which plays a crucial role in planetary atmospheres and influences also dynamics of their interiors. At the same time, crystallography and high-pressure behavior of CO$_2$ are nontrivial and attracted a lot of attention in the last 15 years leaving some topics still unresolved (see reviews \cite{Santoro-Gorelli-CO2-review, Yoo-CO2-review-1, Oganov-C-review}). The exact structure of (pseudo-sixfold) phase-VI (see Refs. \cite{Iota, Montoya, Lee, Sun, Togo}), intermediate character and possible presence of bent molecules in phases II, IV and III - see \cite{Yoo-1, Yoo-2, Iota-Yoo, Yoo-Iota-Cynn, Park, Datchi-IV, Gorelli, Holm, Bonev, Olijnyk-Jephcoat, Yoo-CO2-review-2, Kume}, for example, are still matter of debate. The high-$P$-$T$ regime of CO$_2$ is also disputed as far as several experiments and theoretical works often led to conflicting results \cite{Tschauner, Oganov-CO2, Litasov, Seto, Takafuji, Yoo-i-CO2, Boates-CO2-LLT-1, Boates-CO2-LLT-2, Teweldeberhan}. A liquid-liquid-solid triple point was recently proposed to exist inside the Earth's geotherm region as well \cite{Boates-CO2-LLT-1, Boates-CO2-LLT-2}.

The lowest pressure solid molecular phase of CO$_2$, present on surface of icy caps of Mars, is known as dry ice - phase-I. This quadrupolar $Pa3$ structure transforms between 12-18 GPa into $Cmca$ phase-III \cite{Aoki, Kuchta-Etters, Yoo-1} with molecules aligned in planes, which on further compression transforms into tetrahedral phase-V \cite{Iota-Yoo-Cynn, Yoo-1}, recently identified as $\beta$-cristobalite $I\bar42d$ structure \cite{Datchi-V, Santoro-3} (though existence of a tridymite-like $P2_12_12_1$ structure at similar conditions is was also proposed \cite{Yoo-2V}). Other stable phases of CO$_2$ include molecular phases II \cite{Iota-Yoo}, IV \cite{Yoo-Iota-Cynn, Datchi-IV} and VII \cite{Giordano-Datchi}, polymeric phase-VI \cite{Iota} and other newly discovered forms - possibly polymeric phase-VIII \cite{Sengupta-1}, two tetrahedral structures of coesite-I (phase-IXa) and coesite-II (phase-IXb) \cite{Sengupta-2} and ionic phase $i$-CO$_2$ \cite{Yoo-i-CO2}.

A specific property of phase-V (and all tetrahedral structures of CO$_2$) is extreme rigidity of intertetrahedral C-O-C angle that is represented by energy calculations of $I\bar42d$ phase \cite{Dong-1} and H$_6$C$_2$O$_7$ molecule \cite{Oganov-CO2}, which both show a dramatic increase in energy with variation of the angle out of deep minimum placed near 125$\degree$. This behavior is in a sharp contrast with SiO$_2$, where the minimum is shallow and allows silica to form a rich variety of $sp^3$-polymorphs \cite{Dong-1} unlike the situation in CO$_2$. High stability of tetrahedral over possible octahedral structures in CO$_2$ is also obvious and might be connected to small size of carbon atoms that allow them to occupy interstitial sites of the close-packed oxygen sublattice \cite{Lee}. Stiff C-O-C angle is directly connected also to low compressibility of tetrahedral CO$_2$.

As far as double C=O bond is one of the most stable chemical bonds, molecules sustain large overpressurization before they break and initiate transformation into a single-bonded network. Molecular phase-III hence persists to (60 GPa, 300 K) and to (40 GPa, 1800 K) \cite{Yoo-1}, though the equilibrium transition pressure is according to recent experiment \cite{Sengupta-3} and enthalpy calculations \cite{Tschauner, Togo, Oganov-CO2} only around 20 GPa. The molecular-to-nonmolecular transition is therefore associated with high free energy barriers that lead to negative slope of (kinetic) transition line \cite{Santoro-Gorelli} and possibly also to amorphization at low and moderate temperatures when system is not able to complete the transition and remains stuck in a disordered state.

SiO$_2$ and GeO$_2$ are archetypal glass-forming materials exhibiting low and high density tetrahedral amorphous forms (LDA and HDA) as well as octahedral forms and forms containing fivefold coordinations \cite{Itie-GeO2, Guthrie-GeO2, Hong-GeO2, Lelong-GeO2}. The first prediction of tetrahedra-based amorphous CO$_2$ was based on $ab$ $initio$ molecular dynamics (MD) simulations in the work of Serra $et$ $al.$ \cite{Serra} in 1999 and the first observation of amorphous CO$_2$ was reported two years later \cite{Yoo-Iota-Cynn}. It was suggested from the Raman spectra that the extended amorphous solid is formed by a mixture of three-coordinated ($3c$) and four-coordinated ($4c$) carbon atoms \cite{Yoo-Iota-Cynn}, which would be a novel property amongst the group-IV dioxides. Synthesis of a-CO$_2$ was also reported in Ref. \cite{Santoro-2} and in the further experiments, Santoro $et$ $al.$ \cite{Santoro-1} suggested from Raman spectra that amorphous polymeric form of CO$_2$, named ''a-carbonia'', is a glassy counterpart of phase-V and is also similar to tetrahedral a-silica. Another experimental study was performed at room temperature by Kume $et$ $al.$ \cite{Kume} and a-carbonia was proposed to be related to phase-VI. In these experiments, a-carbonia was decompressed to ambient conditions and a transformation into a molecular amorphous form was observed - at 16 GPa \cite{Santoro-1} and below 30 GPa \cite{Kume}. Amorphization in the higher $P$ region was observed on further compression of phases V, VI and coesite-CO$_2$ over 1 Mbar \cite{Yoo-i-CO2}.

Amorphous CO$_2$ was studied also by first principles simulations (MD \cite{Montoya} and metadynamics \cite{Sun}), which confirmed the picture of mixed three and fourfold nature of a-carbonia, while in both references roughly equal number of $3c$ and $4c$ atoms was reported. Recent $ab$ $initio$ calculations proposed also existence of a first-order LLT between molecular and polymeric liquids \cite{Boates-CO2-LLT-1, Boates-CO2-LLT-2} - the polymeric liquid is formed from the molecular one starting as predominantly $3c$ but evolves upon compression to $4c$-dominated liquid form.

The experimental and theoretical works leave several questions about amorphous carbon dioxide open. In particular, what is the actual structure of a-carbonia - what is the stable ratio of $3c$ and $4c$ carbons (3-4 ratio) at different pressures and what is the structural relation between a-carbonia and crystalline CO$_2$. Furthermore, how the structural evolution from a-carbonia to the molecular amorphous form proceeds and what exactly happens upon (de)compression - if the transformations are continuous or discontinuous and whether molecules can eventually coexist with the polymeric form. In this paper, we aim at resolving these questions using \textit{ab initio} MD.

The paper is organized as follows. In part II, we describe methods and main findings of our simulations that are analyzed in part III. The analysis includes investigation of structural properties of polymeric amorphous forms and mixed molecular-polymeric form and their possible relations to crystalline phases. Next, mechanical stability of the observed forms is analyzed and compared. Finally, enthalpies, compressibilities, electronic properties and structure factors of all forms are calculated and discussed. Finally, we summarize our observations and suggestions for further study in the conclusions.

\section{Simulation methods and protocol}

We used standard density functional theory (DFT) based codes VASP 4 and 5 \cite{VASP-1, VASP-2}. To simulate systems under constant pressure with VASP 4 version, we employed slightly modified Berendsen algorithm \cite{Berendsen} where cell parameters and atomic positions and velocities were rescaled according to difference of external and internal stress tensor every 20 MD timesteps (more details are described in Ref. \cite{Plasienka-Martonak-1}). Simulations performed with newer versions of VASP 5 were carried out with the implemented Parrinello-Rahman (PR) barostat \cite{PR-1} working together with the Langevin thermostat generating the $NPT$ ensemble. PAW pseudopotentials and the PBE functional \cite{PBE} were used to describe four/six valence electrons for each carbon/oxygen atom, using energy cutoff 450 eV and $\Gamma$-point sampling of the Brillouin zone.

\begin{figure}[h]
\includegraphics[width=\columnwidth]{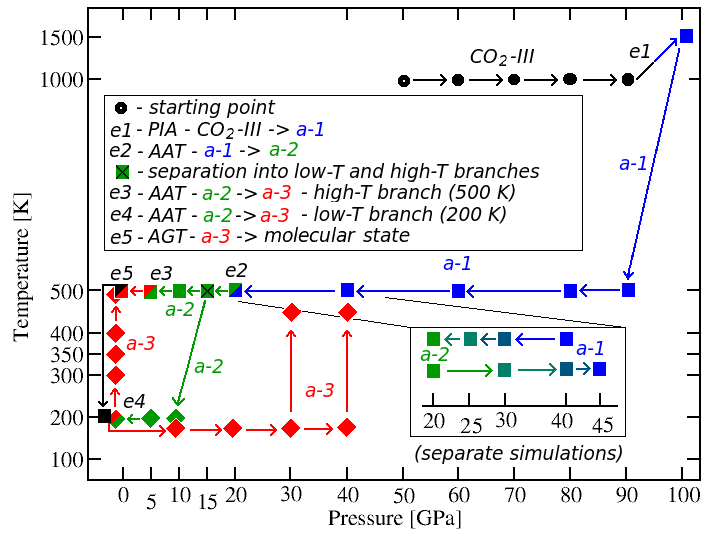}
\caption{(Color online) Simulation protocol of solid CO$_2$. Starting from phase-III (black dots and arrows) at 50 GPa and 1000 K, PIA was observed at 100 GPa - event-1 (e1) leading to formation of polymeric a-carbonia form denoted as \textit{a-1} (blue squares and arrows). Decompression at 500 K led to creation of another version of polymeric a-carbonia \textit{a-2} (green squares) - event-2 (e2) and on further pressure drop \textit{a-3} (red squares) was formed at 5 GPa - event e3. The same \textit{a-3} form (red diamonds) was created independently at 0 GPa and 200 K - e4. In the 500 K simulation branch, molecular state (black) appeared at 0 GPa - e5. In the inset are shown separate simulations of compression of \textit{a-2} and decompression of \textit{a-1} between 20 and 45 GPa at 500 K performed in more pressure steps and for longer times as in the original simulation. The turquoise color of the squares represents forms with intermediate character between blue \textit{a-1} and green \textit{a-2}.}
\label{fig:protocol}
\end{figure}

We ran all simulations on fairly large systems consisting of 108 CO$_2$ molecules. After optimization of phase-III (black) to 50 GPa, we started dynamical simulations by heating the system to 1000 K and then increasing pressure in 10 GPa steps (Fig.~\ref{fig:protocol}). Sample amorphized upon compression from (90 GPa, 1000 K) to (100 GPa, 1500 K) similarly to previous DFT simulations \cite{Serra}. The resulting polymeric a-carbonia form (blue), which we denote as \textit{a-1} here, was dominated by $4c$ carbons (CO$_4$ tetrahedras). From this point, we started decompression at temperature of 500 K in order to study the evolution of the amorphous state. After bringing system to 20 GPa, \textit{a-1} was transformed to a different polymeric form with similar proportion of $3c$ and $4c$ carbons - \textit{a-2} (green). Amorphous forms \textit{a-1} and \textit{a-2} appear in our simulations as two limiting (high and low pressure) realizations of polymeric a-carbonia because they transform into each other gradually between 20 and 45 GPa as observed in separate calculations shown in the Fig.~\ref{fig:protocol} inset.

Afterwards, we proceeded with decompression along two separate pathways - at 500 K and at 200 K \footnote{Simulations of initial compression and 500 K branch decompression were performed with VASP 4 and Berendsen barostat, while lower-$T$ branch decompression and all other calculations were calculated with VASP 5 and PR barostat.}. In the 500 K branch we observed a formation of mixed molecular-polymeric form - \textit{a-3} (red) at 5 GPa, while the same form appeared also in the lower 200 K branch at 0 GPa. The two independent kinetic pathways leading to the same amorphous form - with equal proportion of coordinations, suggest that \textit{a-3} form is not a mere artifact of the simulation timescale, but instead a form with pronounced metastability. This is further supported by stability of \textit{a-3} form obtained at (0 GPa, 200 K) on its subsequent compression to 40 GPa and 500 K (see Fig.~\ref{fig:protocol}). In the 500 K branch, \textit{a-3} completely depolymerized into a molecular state (black), which behaved like gas at the simulation temperature of 500 K (therefore denoted as amorphous-gaseous transition - AGT on Fig.~\ref{fig:protocol}) and solidified to molecular amorphous form at 200 K. The total simulation time of our 324 atomic system exceeds 1 ns.

\section{Results and discussion}

\begin{figure}[h]
\includegraphics[width=\columnwidth]{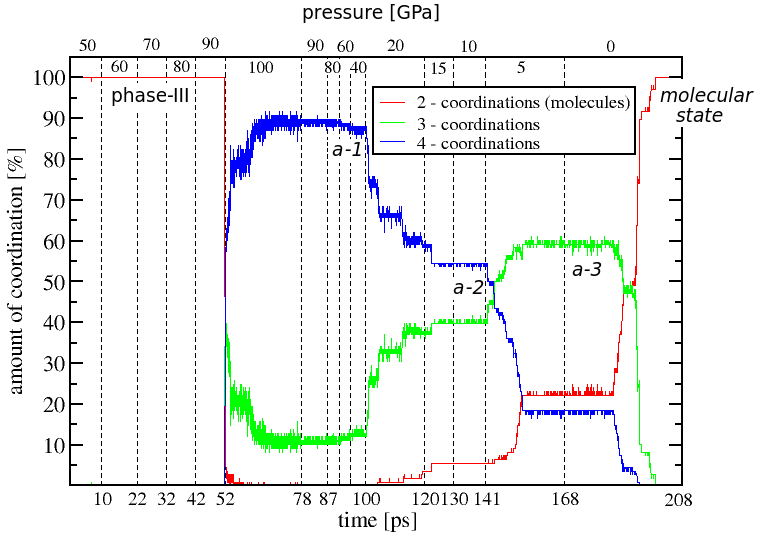}
\caption{(Color online) Amount of carbon coordinations [\%] on compression (to 78 ps), where \textit{a-1} was formed, and on higher-$T$ branch (500 K) decompression (from 78 ps to 208 ps), where all other forms appeared. The CO$_2$ molecules are shown in red (starting and ending at 100\%), $3c$ are in green and $4c$ in blue ($4c$-curve is systematically larger than $3c$-curve from 52 to 144 ps and lower thereafter). The coordination limit was 1.7 \AA. Vertical dashed lines divide graph into regions of different simulation pressures that are labeled on top and are changed at corresponding times labeled at bottom.}
\label{fig:coordinations}
\end{figure}

To characterize the various amorphous forms, we analyzed proportion of different carbon coordinations. On Fig.~\ref{fig:coordinations}, we show the entire run of initial compression and 500 K branch decompression from Fig.~\ref{fig:protocol} spanning total simulation time of 208 ps, where all amorphous states were observed. The amount of carbon two (CO$_2$ molecules), three and four coordinations are shown in red, green and blue, respectively.

Compression from 90 to 100 GPa (starting at 52 ps) caused immediate breakdown of all molecules and formation of fully-extended disordered network - \textit{a-1}, with 88\% of $4c$ carbons and remaining 12\% of $3c$ atoms. The form \textit{a-1} persisted unchanged on decompression to 40 GPa, but at 20 GPa (at 100 ps) a number of bonds desaturated and a new mixed $3c$ (40\%) and $4c$ (55\%) form - \textit{a-2} containing also 5\% of molecules appeared. During simulations at 5 GPa, this form transformed into \textit{a-3} (at about 155 ps) consisting of 18\% of $4c$ carbons, 60\% of 3c carbons and 22\% of molecules (the same as in the lower 200 K branch, not shown here). At 0 GPa and 500 K, \textit{a-3} finally started to decay (at 185 ps) and all molecules were recovered shortafter (at 199 ps).

Separate simulations of compression of \textit{a-2} from 10 to 45 GPa and decompression of \textit{a-1} from 40 to 20 GPa at 500 K (see inset of Fig.~\ref{fig:protocol}) revealed that both forms gradually transform into each other in this pressure window, which can be viewed as a continuous transformation between the two limiting states of amorphous a-carbonia - the high-pressure tetrahedral form \textit{a-1} and the low-pressure form \textit{a-2}. The forms are limiting forms of polymeric a-carbonia in the sense that further compression of \textit{a-1} does not induce further structural transformation (bonds are saturated) and decompression of \textit{a-2} leads directly into a different amorphous regime represented by the mixed molecular-polymeric form \textit{a-3} (and not to a different polymeric state).

\subsection{Structure of nonmolecular a-carbonia}

Amorphization occurs very fast upon the compression to 100 GPa and is accompanied by a large volume collapse and complete structural reorganization. From the distribution of intramolecular O=C=O angles, we observed that molecules always remained linear lacking any systematic bending before the onset of amorphization.

\begin{figure}[h]
\begin{tabular}{cc}
\fbox{\includegraphics[width=0.8\columnwidth]{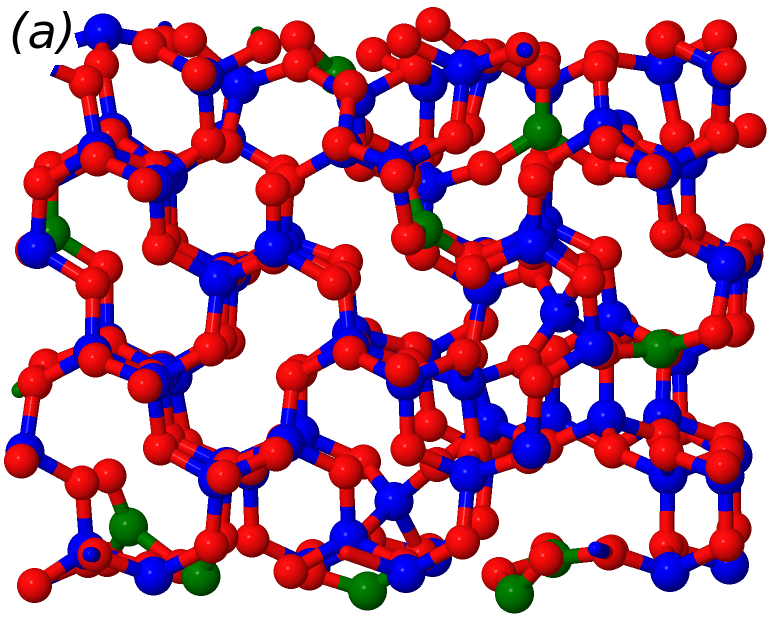}}\\
\fbox{\includegraphics[width=0.8\columnwidth]{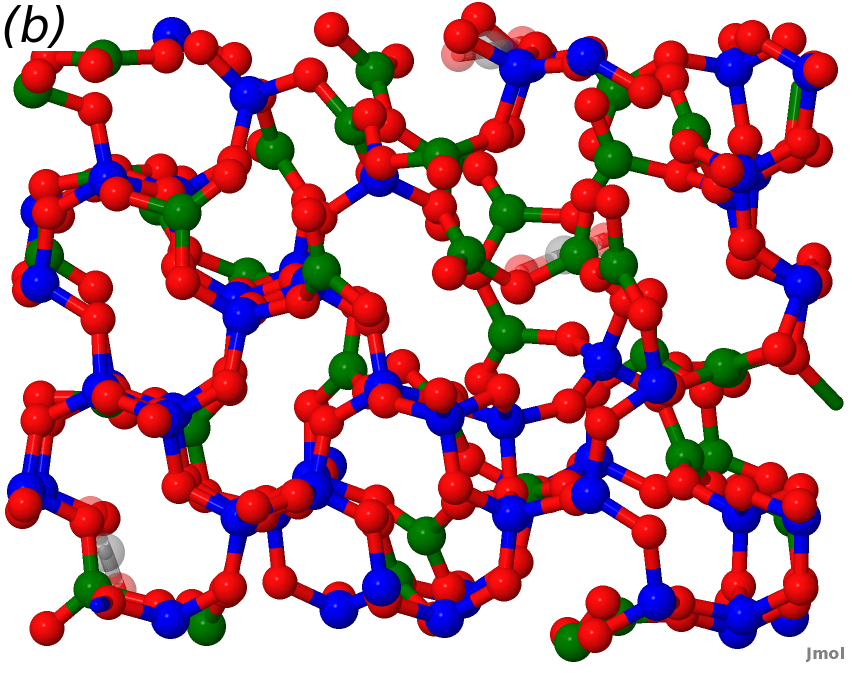}}\\
\end{tabular}
\caption{(Color online) (a) Snapshot of \textit{a-1} at 40 GPa and (b) of \textit{a-2} at 20 GPa and 500 K. $4c$ and $3c$ carbon atoms are marked as blue and green spheres, respectively. Pictures were generated by Jmol \cite{Jmol}.}
\label{fig:phase1}
\end{figure}

\begin{figure}[h]
\includegraphics[width=0.85\columnwidth]{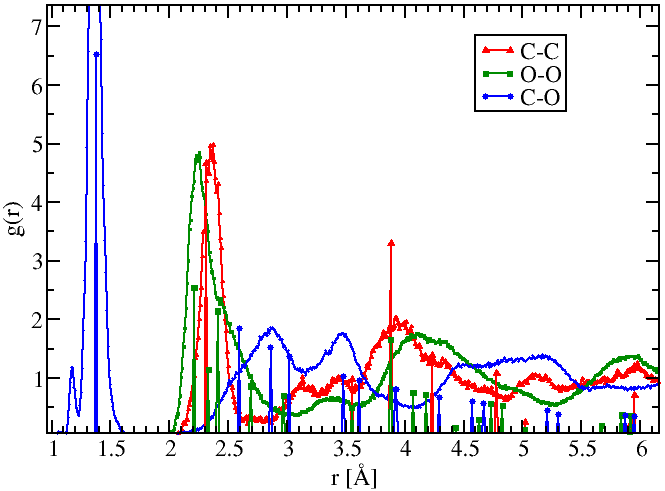}
\includegraphics[width=0.85\columnwidth]{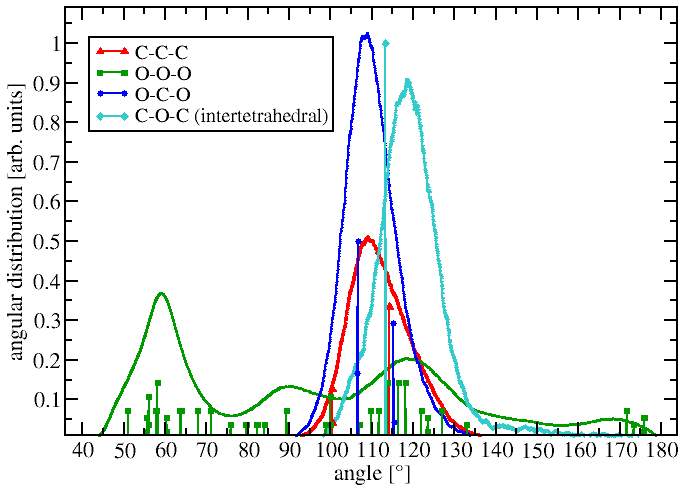}
\caption{(Color online) (Upper panel) RDFs of type C-C (red triangles), O-O (green squares) and C-O (blue circles) for \textit{a-1} (broad peaks) at 40 GPa, 500 K and zero-$T$ structure of phase-V at 41 GPa from data provided by Datchi $et$ $al.$ \cite{Datchi-V} (sharp peaks). The tiny first C-O peak represents C=O bonds from small number of $3c$ carbons. (Lower panel) ADFs of type C-C-C (red triangles), O-O-O (green squares), O-C-O (blue circles) and intertetrahedral C-O-C (turquoise diamonds) - all calculated within the first RDF minima of \textit{a-1}.}
\label{fig:RDF-ADF}
\end{figure}

Structure of \textit{a-1} form is shown on Fig.~\ref{fig:phase1}(a), where a nanocrystallite \footnote{In the experiments, nanocrystalline and amorphous phases also cannot be always distinguished due to finite resolution ability of the device.} of phase-V seems to be formed inside the amorphous network. This points to a structural correspondence between tetrahedral-like form of a-carbonia - \textit{a-1} and crystalline phase-V. To prove their relation, short-range order of both forms was investigated and depicted onto Fig.~\ref{fig:RDF-ADF}. Radial distribution functions (RDFs) - upper panel and angular distribution functions (ADFs) - lower panel of \textit{a-1} and phase-V are shown along each other, while all distribution peaks of \textit{a-1} are broad and for phase-V are sharp. One can see from the figures clearly that all broad peaks of \textit{a-1} well cover the corresponding sharp peaks of phase-V. Regarding the nearest neighbors, value of the C-C coordination number $N^{CC}_\text{C}$ of \textit{a-1} is 3.83 at cutoff 2.6 \AA \,and $N^{OO}_\text{C}$ = 11.87 at 2.7 \AA. The C$\rightarrow$O coordination $N^{C\rightarrow O}_\text{C}$ = 3.88 and the O$\rightarrow$C coordination $N^{O\rightarrow C}_\text{C}$ = 1.94 at cutoff 1.7 \AA. The corresponding coordination numbers for phase-V are similar - 4, 12, 4 and 2, respectively at the same cutoff values. The \textit{a-1} form can be therefore regarded as an amorphous version of phase-V, as was suggested for experimentally observed a-carbonia from its Raman spectra \cite{Santoro-1}.

An important property of phase-V is rigidity of the intertetrahedral C-O-C angle. The peak of the C-O-C ADF in \textit{a-1} is placed around 118$\degree$ - Fig.~\ref{fig:RDF-ADF} lower panel turquoise curve. This is in a good agreement with the calculated ideal value - 125-130$\degree$ \cite{Dong-1} or 124$\degree$ \cite{Oganov-CO2}) and also with the measured angle - 113.2$\degree$ \cite{Datchi-V}. The distribution is, however, quite sharp for an amorphous solid (compare e.g. with the wide Ge-O-Ge distribution in a-germania \cite{Peralta-GeO2}) and indicates that stiffness of the C-O-C angle is a basic property of CO$_2$ that is inherited into the amorphous regime.

Similar amorphous form as is our predominantly tetrahedral \textit{a-1} was reported in the previous MD simulations \cite{Serra}. In the original study, it was described as a tetrahedral amorphous solid, while in the subsequent work \cite{Montoya}, it was stated that it contained unspecified number of unsaturated bonds (we remind \textit{a-1} also contains 12\% of $3c$ carbons). A glass with similar structure was obtained by quenching from $4c$-dominated liquid state \cite{Boates-CO2-LLT-2}.

The second limiting a-carbonia form \textit{a-2} - Fig.~\ref{fig:phase1}(b) contains only slightly higher number of $4c$ than $3c$ carbons. Very similar forms like this were obtained in the previous $ab$ $initio$ simulations, which were performed along different $P$-$T$ pathways \cite{Montoya, Sun}. Direct experimental evidence about the quantities of carbon coordinations (as determined e.g. for amorphous GeO$_2$ \cite{Lelong-GeO2}) is as yet not available.

\subsection{Structure of molecular-polymeric form}

\begin{figure}[h]
\fbox{\includegraphics[width=0.8\columnwidth]{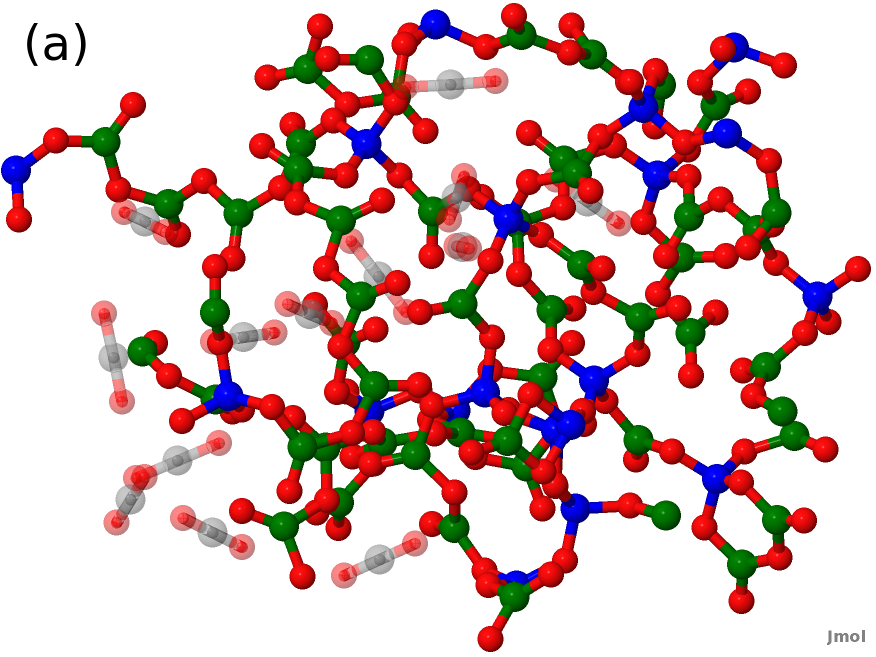}}\\
\begin{tabular}{cc}
\includegraphics[width=0.35\columnwidth]{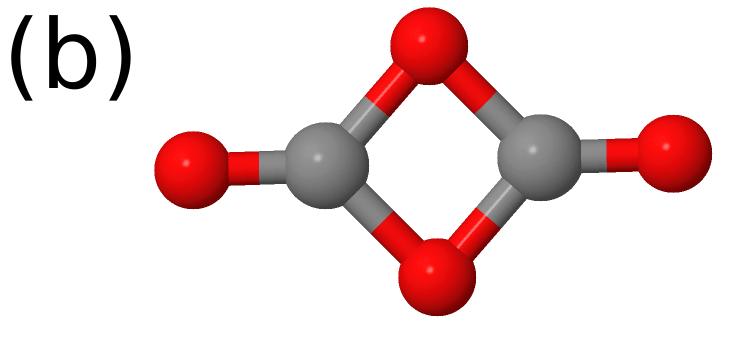}&\hspace{1cm}
\includegraphics[width=0.3\columnwidth]{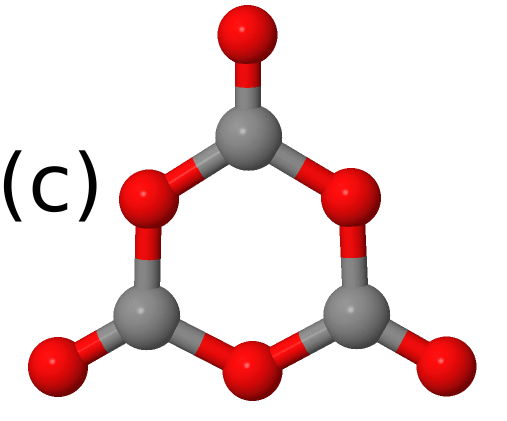}\\
\end{tabular}
\caption{(Color online) (a) Amorphous form \textit{a-3} at 5 GPa and 500 K. One closed chain of two $3c$ atoms pinned to one $4c$ carbon is placed near the down right edge of the picture. $4c$ and $3c$ carbons are distinguished as blue and green atoms, isolated CO$_2$ molecules are marked translucent. (b) C$_2$O$_4$ dimer and (c) C$_3$O$_6$ trimer occurring in the \textit{a-3} form.}
\label{fig:phase-3}
\end{figure}

At two different $P$-$T$ points (e3 and e4 on Fig.~\ref{fig:protocol}), the \textit{a-2} form independently transformed into substantially different molecular-polymeric amorphous state \textit{a-3} shown on Fig.~\ref{fig:phase-3}(a). To or knowledge this form of a-CO$_2$ was not discussed so far. The \textit{a-3} form represents a mechanically stable local packing of CO$_2$, CO$_3$ and CO$_4$ units forming a bridge between molecular and polymeric amorphous states. The form is composed of three basic units - $sp^3$-tetrahedras CO$_4$, triangles CO$_3$ and linear molecules CO$_2$ \footnote{C-O type RDF of \textit{a-3} shows two distinct nearest peaks corresponding to two different C=O double bonds present in it - shorter 1.17 \AA \,in CO$_2$ molecules and longer 1.18 \AA \,coming from double bonds on $3c$ carbons.}. The molecules remain isolated, while $3c$ and $4c$ carbons connect and form several structural patterns. The most abundant are polymeric chains formed by series of $3c$ carbons pinned to two different $4c$ carbons (nodes), which are present in various lengths and torsions. Next, we observe closed chains beginning and ending in the same $4c$ carbon and occasionally also entirely three-coordinated loops (cyclic molecules) shown on Fig.~\ref{fig:phase-3}(b) and (c). Some of the cyclic (CO$_2$)$_x$ oligomers were already studied by methods of theoretical chemistry \cite{Lewars, Frapper-Saillard}.

Formation of C$_2$O$_4$ dimeric molecules were observed in MD of high-$T$ liquid phases \cite{Tassone, Boates-CO2-LLT-2}. In Ref. \cite{Tassone}, for example, metastability of dimers, which we indeed observe in \textit{a-3}, was proposed to take place at low temperature. Moreover, it was also suggested that existence of these dimers may represent a kinetic intermediate step on the transformation to some three-coordinated crystalline phase, which is discussed in the next paragraph.
In another theoretical calculations of liquid CO$_2$, presence of unstable CO$_2$ molecules in predominantly polymeric liquid form near the proposed LLT line region \cite{Boates-CO2-LLT-2} was also observed. This indicates that mechanically stable mixture of molecular and nonmolecular state at low temperatures (in solid state), where the kinetics is considerably slower, may be possible.


We now briefly discuss possible thermodynamical background of amorphous forms containing $3c$ carbons. Presence of these $3c$ carbon atoms in a-carbonia and also in form \textit{a-3} opens a natural question whether $3c$ atoms can form some stable or at least metastable structure. While no such phase has been observed experimentally, some theoretical hints exist \cite{Tassone, Frapper-Saillard, Montoya}. In our case, fact that \textit{a-3} contains chains of $3c$ atoms as basic building blocks points to the possible relation to a hypothetical crystalline phase composed of infinite parallel chains of $3c$ carbon atoms.
Possibility of such chain forms has been proposed in some previous studies \cite{Tassone, Frapper-Saillard, Montoya}

\begin{figure}[h]
\begin{tabular}{cc}
\includegraphics[width=0.65\columnwidth]{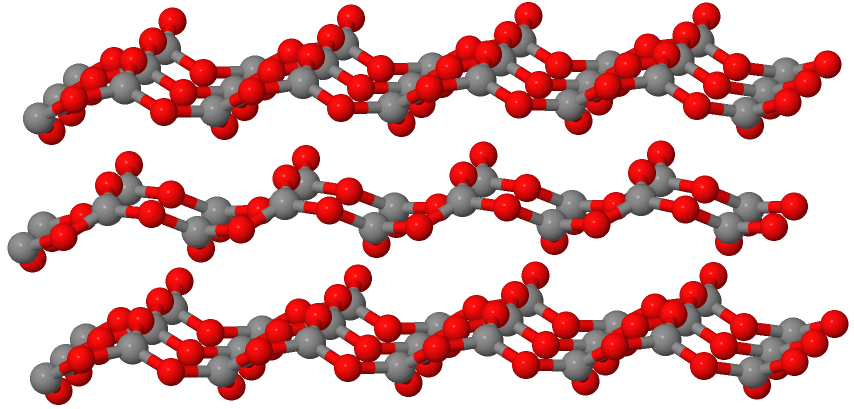}&
\includegraphics[width=0.35\columnwidth]{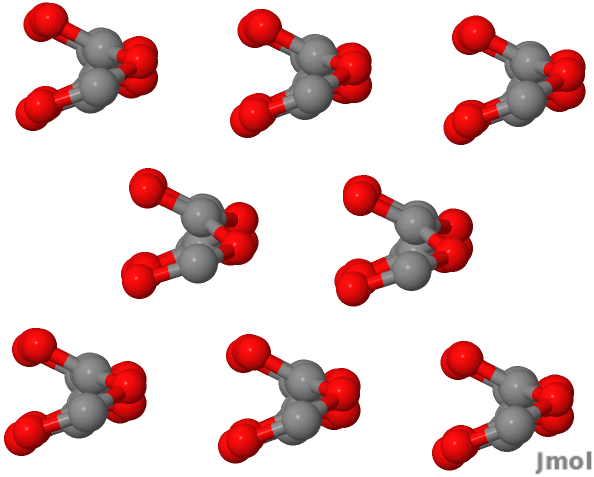}\\
\end{tabular}
\caption{(Color online) Phase-\textit{3C} at 10 GPa - front view (left) and chain axis direction side view (right).}
\label{fig:phase3C}
\end{figure}

The initial guess in our search for the $3c$ form was inspired by Ref. \cite{Montoya} and picture of the structure depicted on Fig.~4(b) therein. Following optimizations at several pressures, we found a structure denoted as phase-\textit{3C}, which is formed by linear zig-zag chains aligned in mutually shifted planes (Fig.~\ref{fig:phase3C}). Phase-\textit{3C} has lower enthalpy than molecular phases at pressures over 40 GPa (see Fig.~\ref{fig:enthalpies}). Phase-\textit{3C} was stable in dynamical simulations at 0 GPa and 200 K for several tens of picoseconds.

\subsection{Structural stability}

\begin{figure}
\includegraphics[width=\columnwidth]{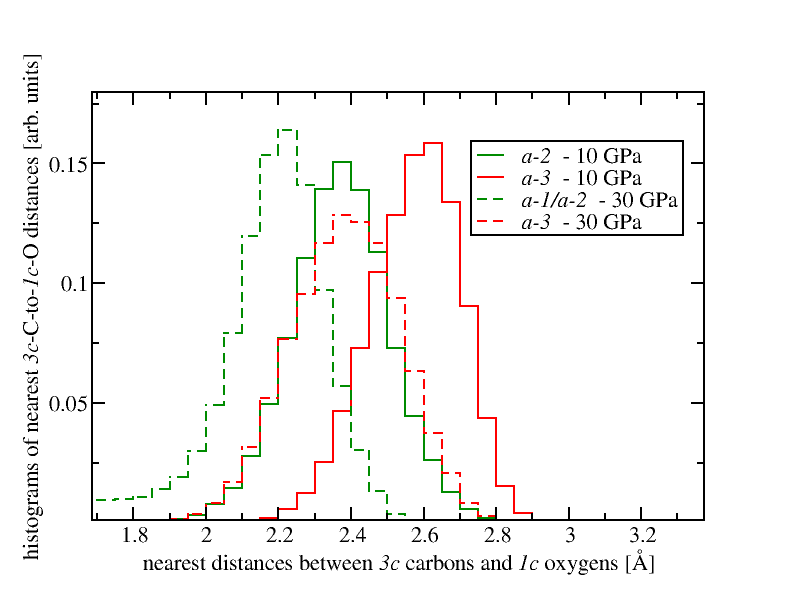}
\includegraphics[width=\columnwidth]{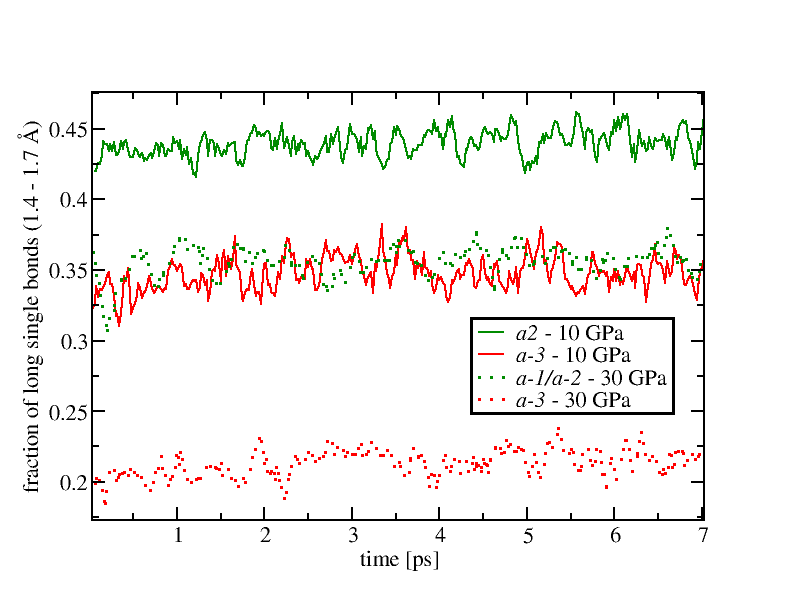}
\caption{(Color online) (a) Histograms of nearest distances between
  $3c$ carbons and $1c$ oxygens for polymeric a-carbonia \textit{a-2}
  (green histogram) at 10 GPa and intermediate state \textit{a-1}/\textit{a-2}
  (dashed green) at 30 GPa and \textit{a-3} at 10 GPa (red) and at 30
  GPa (red dashed), all at 500 K. The histograms are scaled to the
  total number of $3c$ carbons in the system. (b) Fraction of C-O
  single bonds with length between 1.4 and 1.7 \AA \, calculated for
  each frame during a 7 ps time interval extracted from the MD runs
  and scaled to total numbers of C-O single bonds. Colors and line
  styles represent the same systems and conditions as in (a) -
  systems at 30 GPa are marked with dotted lines. Full lines and
  dashed/dotted lines are to be compared separately as different line
  styles represent different simulation pressures and line colors distinct between
  polymeric and \textit{a-3} form.}
\label{fig:a3-stability}
\end{figure}

We turn now to discussion of the behavior of the observed amorphous
forms upon change of pressure, namely gradual transformation between
limiting polymeric forms \textit{a-1} and \textit{a-2} and pronounced
metastability of the molecular-polymeric \textit{a-3} form. To relate
the stability upon compression to the network structure we analyze the
distribution of nearest distances between possible reaction sites,
namely $3c$ carbons and single-coordinated ($1c$) oxygen atoms (with
double bonds). The distribution of the nearest distances at certain
conditions thus reflects the potential ability to turn $3c$ carbons into
$4c$ ones upon compression. The respective histograms are shown in
Fig.~\ref{fig:a3-stability}(a) where we compare at 10 GPa \textit{a-3}
and \textit{a-2} and at 30 GPa \textit{a-3} and intermediate state
between \textit{a-1} and \textit{a-2} (containing 63\% of $4c$
carbons). At both pressures of 10 GPa and 30 GPa, histograms of
\textit{a-3} are shifted away from the corresponding polymeric ones
indicating that less possible reaction sites of $3c \rightarrow 4c$ C
transitions are available. Moreover, at 30 GPa, the \textit{a-1}/\textit{a-2}
state contains sizable amount of nearest $3c$-C-to-$1c$-O distances
between 1.7 and 2 \AA, which allows for gradual evolution of coordinations
in the polymeric regime between the limiting forms \textit{a-1} and \textit{a-2}.

The structural stability of the system during decompression, on the other
hand, can be related to the relative number of most strained single C-O bonds
which are prone to break during the volume increase. To this end we
count the fraction of elongated single bonds in the length interval
from 1.4 to 1.7 \AA \,- Fig.~\ref{fig:a3-stability}(b). It can be
clearly seen that the \textit{a-3} form compared to polymeric forms
contains lower amount of elongated single bonds and thus is more
stable also with respect to decompression.

\subsection{Thermodynamical, mechanical, electronic and structural properties}

\begin{figure}
\includegraphics[width=\columnwidth]{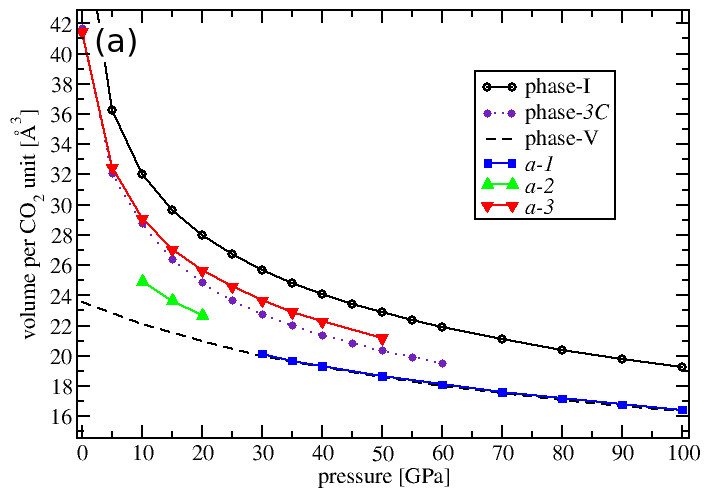}
\includegraphics[width=\columnwidth]{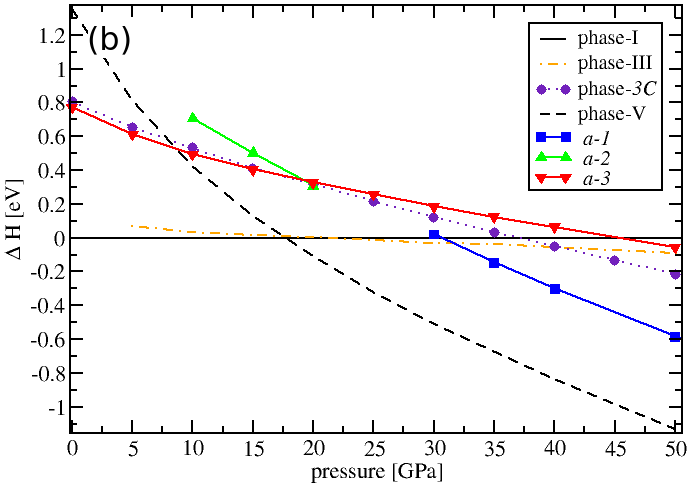}
\caption{(Color online) (a) Equation of states for V(P) - volume per CO$_2$ unit versus pressure of phase-I (line with empty circles), phase-V (dashed line), phase-\textit{3C} (violet dotted line with circles) and amorphous \textit{a-1} (blue squares), \textit{a-2} (green up triangles) and \textit{a-3} (red down triangles) in 1 Mbar range. (b) Enthalpies from 0 to 50 GPa relative to phase-I (horizontal line) per CO$_2$ unit of crystalline phase-III (orange dashed-dotted line), phase-\textit{3C} (violet dotted line with circles), phase-V (black dashed line) and amorphous \textit{a-1} (blue squares), \textit{a-2} (green up triangles) and \textit{a-3} (red down triangles). All phase curves are shown only in their stability regions, where they survived optimization.}
\label{fig:enthalpies}
\end{figure}

To analyze relative zero-$T$ stability of the discussed amorphous and
crystalline forms, we calculated equations of states for volume and enthalpy
versus pressure - Fig.~\ref{fig:enthalpies}(a) and (b), respectively.
Our calculated enthalpy functions show that phase-V becomes more stable
than phase-III over 17.5 GPa and molecular phase-I crosses phase-III curve
at around 20 GPa, similarly to Ref. \cite{Bonev}. The three-coordinated
Phase-\textit{3C} (stable up to 60 GPa) is favored over phase-V below 8 GPa and over phase-III
above 40 GPa, though it is metastable at all pressures.

Regarding the amorphous forms, \textit{a-1} curve systematically copies
curve of the crystalline phase-V with about 0.7 eV higher enthalpy values.
The \textit{a-2} form survived optimization only between 10 and 20 GPa, while
\textit{a-1} decayed below 30 GPa. This nonstability of both limiting
polymeric forms between 20 and 30 GPa observed in the optimization process
is related to the existence of the pressure window, where intermediate states between \textit{a-1} and
\textit{a-2} were observed in the MD simulations (see inset of Fig.~\ref{fig:protocol}).
Form \textit{a-3} was stable between 0 and 50 GPa and its enthalpy shows that is more
stable than phase-V below 8.5 GPa. This analysis shows that in certain
pressure intervals, the amorphous forms have lower enthalpies compared
to some competing crystalline phases and might synthesize at carefully
chosen experimental $P$-$T$ conditions.

To calculate bulk moduli of the investigated forms, we run simulations with the Parrinello-Rahman barostat to obtain volume fluctuations $\left< \delta V^2 \right>$ for $NPT$ ensemble. These fluctuations are proportional to the isothermal bulk modulus $B$ according to the fluctuation formula \cite{Allen-Tildesley}
\begin{equation*}
B = \frac{kT}{V} \left< \delta V^2 \right>_{NPT},
\end{equation*}
where $V$ is average volume.

All calculated values of $B$ were extracted from separate MD trajectories at 200 K lasting 40-60 ps to assure converged values of $\left< \delta V^2 \right>_{NPT}$. From Table \ref{tab:bulk} we see that upon progressive transformation from \textit{a-1} to molecular states the value of $B$ decreases by more than an order of magnitude. The \textit{molecular amorphous} corresponds to 200 K quenched molecular amorphous state (glass) obtained from molecular gas that appeared at the end of the original 500 K decompression (see Fig.~\ref{fig:protocol}).

\begin{center}
\begin{table}
\begin{tabular}{|l|c|c|}
\hline
Phase & bulk $B$ [GPa] & $P$ [GPa] \\ \hline
phase-V & 294 & 40 \\ \hline
\textit{a-1} & 282 & 40 \\ \hline
\textit{a-2} & 75 & 5 \\ \hline
\textit{a-3} & 46 & 0 \\ \hline
phase-I & 25 & 0 \\ \hline
\textit{molecular amorphous} & 20 & 0 \\ \hline
\end{tabular}
\caption{Calculated values of $B$ at 200 K from hardest phase-V to soft molecular phases. $B$ of phase-V at 40 GPa is reaching 300 GPa corresponding very well with theoretical predictions - Dong \textit{et al.} \cite{Dong-2, Dong-1} and actual recent experimental measurements - Datchi \textit{et al.} \cite{Datchi-V}.}
\label{tab:bulk}
\end{table}
\end{center}

We also calculated electronic properties of the amorphous forms within the PBE approximation and found that \textit{a-3} is an insulator with energy bandgap 3.35 eV at 5 GPa and \textit{a-2} is a semiconductor with 1.71 eV bandgap at 20 GPa. In \textit{a-1} at 40 GPa, the gap energy is decreased to 1.48 eV, while at 90 GPa it narrows to 0.7 eV. Therefore, polymerization into a tetrahedral-like amorphous form is not followed by metalization, though closure of the bandgap can be expected at Mbar conditions \cite{Yoo-CO2-review-1}. We remark that the predicted LLT in CO$_2$ is also not accompanied by metalization \cite{Boates-CO2-LLT-2}, which was noted to be exceptional for a molecular-polymeric transition in a high-$T$ liquid state.

\begin{figure}
\includegraphics[width=\columnwidth]{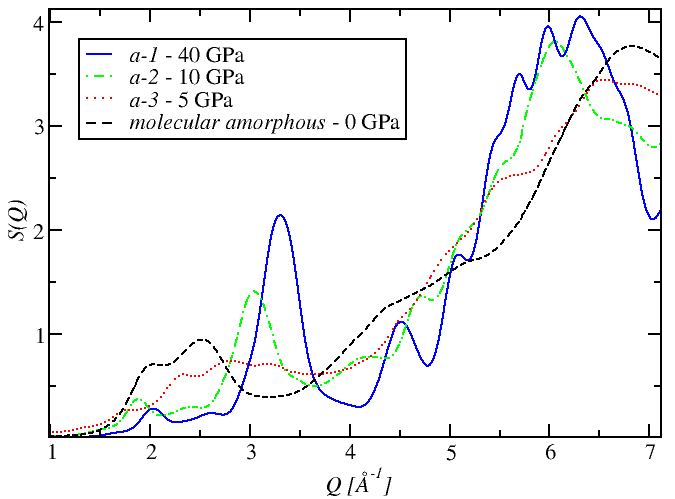}
\caption{(Color online) Static structure factors of \textit{a-1} (blue line) at 40 GPa, \textit{a-2} (green dashed-dotted line) at 10 GPa, \textit{a-3} (red dotted) at 5 GPa and \textit{molecular amorphous} form - glass (dashed black) at 0 GPa. The structural evolution from \textit{a-1} to \textit{a-2}, \textit{a-3} and to \textit{molecular amorphous} form is represented mainly by broadening and shifting of the first main $S(Q)$ peak to lower $Q$ - from 3.3 \AA$^{-1}$ \,in \textit{a-1} to 2.5 \AA$^{-1}$ \,in the molecular state.}
\label{fig:SQ}
\end{figure}

In order to present quantities directly comparable to experiments, we calculated static structure factors $S(Q)$ of the amorphous forms - Fig.~\ref{fig:SQ}.
$S(Q)$ functions were calculated from the MD trajectories by the method described in Refs. \cite{Holender-Morgan, Liang-PhD}. We first calculated $S(\vec{Q}_{hkl})$ at discrete set of $\vec{Q}_{hkl}$ vectors (determined by the periodic boundary conditions) and then made convolution with Gaussian of width 0.1 \AA$^{-1}$ to obtain $S(Q)$. 
This quantity was afterwards averaged over the corresponding trajectory. The normalization of the structure factors was chosen according to Ref. \cite{Santoro-1} (and the Supplemental Material therein), where $S(Q)$ was factorized to the molecular formfactor $(f_C(Q)+2 f_O(Q))^2$, where $f_C(Q)$ and $f_O(Q)$ are atomic formfactors of carbon and oxygen, respectively. The calculated structure factors can then be directly compared to the experimental ones on Figure 4 of Ref. \cite{Santoro-1}.

The experimental curve at 41 GPa appears similarly to our curve of \textit{a-1} at 40 GPa - both contain main first peak at around 3.3 \AA$^{-1}$. Upon decompression, our peak progressively shifts to lower $Q$ values - \textit{a-2} and becomes broader when molecules are created - \textit{a-3}. Finally, upon transition to the molecular state (in the glass), the peak appears to become split which is similar to the experimental data at 12 GPa, when a-carbonia was already transformed into the molecular amorphous form. This indicates that our simulated forms of a-carbonia could be associated with the earlier experimental observations \cite{Santoro-1}.

\section{Conclusions}

Using ab initio MD simulations we performed decompression of polymeric
a-carbonia initially prepared at high-pressure and observed several
amorphous forms which behaved like mechanically stable and
long-living metastable states. As the pressure decreases the original
high-pressure polymeric form (\textit{a-1}) with mostly four-coordinated carbon
atoms and tetrahedral geometry first gradually transforms into less
dense structures eventually reaching a limiting form (\textit{a-2}) with
roughly equal number of four and three-coordinated carbons. Both these
forms were observed in the earlier simulations \cite{Serra, Sun, Montoya}.
Upon further decompression molecules start to appear and a
new mixed molecular-polymeric form (\textit{a-3}) is found before the system
finally transforms to a fully molecular state. In the mixed form
four-coordinated carbon atoms act as nodes that are connected by
chains of three-coordinated carbons, while space between the chains is
filled with molecules. Compared to the polymeric forms the mixed a-3
form appears to have pronounced metastability which can be related to
different distribution of certain interatomic distances and bond lengths.
Due to the large gap between the timescale of experiments and
simulations it is not trivial to extrapolate the metastability
observed in our simulations to true metastability at experimental
conditions. However, the facts that the two polymeric forms were also
reported in previous simulations and that the mixed
molecular-polymeric form was prepared in our simulations in two
independent pathways suggest that these states might indeed represent
observable amorphous phases. We believe that it would be interesting
to experimentally verify our predictions by carefully monitoring the
structural evolution in a-CO$_2$ in a slow gradual decompression
performed at low temperature, where the kinetics is slower.

\begin{acknowledgments}
This work was supported by the Slovak Research and Development Agency
under Contract No.~APVV-0558-10 and No.~APVV-0733-11 and by the
project implementation 26220220004 within the Research \& Development
Operational Programme funded by the ERDF. Part of the calculations
were performed in the Computing Centre of the Slovak Academy of
Sciences using the supercomputing infrastructure acquired in project
ITMS 26230120002 and 26210120002 (Slovak infrastructure for
high-performance computing) supported by the Research \& Development
Operational Programme funded by the ERDF.
\end{acknowledgments}

\end{document}